\documentclass[%
 reprint,
 amsmath,amssymb,
aps,
]{revtex4-1}

\usepackage{graphicx}% Include figure files
\usepackage{dcolumn}% Align table columns on decimal point
\usepackage{bm}% bold math
\usepackage{hyperref}% add hypertext capabilities
\usepackage{algorithm}
\begin{document}
%\preprint{APS/123-QED}

\title{Transfer entropy: where Shannon meets Turing}% Force line breaks with \

\author{David Sigtermans\\ david.sigtermans@asml.com}
%\email{david.sigtermans@asml.com}
\affiliation{%
 ASML Netherlands B.V.\
}%

\date{\today}% It is always \today, today,
        % but any date may be explicitly specified

\begin{abstract}
Transfer entropy is capable of capturing nonlinear source-destination relations between multivariate time series. It is a measure of association between source data that are transformed into destination data via a set of linear transformations between their probability mass functions. The resulting tensor formalism is used to show that in specific cases, e.g., in the case the system consists of three stochastic processes, bivariate analysis suffices to distinguish true relations from false relations. This allows us to determine the causal structure as far as encoded in the probability mass functions of noisy data. The tensor formalism was also used to derive the Data Processing Inequality for transfer entropy.
\end{abstract}

\maketitle
Efficient inference of the source-destination relations within a complex system from observational is essentially a ``catch 22'' situation. Pairwise analysis is relatively cheap, but a bivariate analysis will reveal a relation between two non-interacting processes that are correlated only due to a common source. Multivariate analysis leads to a higher precision but it is computational very costly. For transfer entropy \cite{Schreiber} several approaches have been developed to resolve the computational and precision issue, e.g., \cite{Runge,Wollstadt}).

In this letter we report a novel approach. Our vantage point is that of a machine builder. Photolithography machines are extremely complex systems consisting of tens of thousands of interacting components. Designing and building these systems is impossible without the notion of causality. Because the output of a machine can be thought of as the result of a computation, a Turing machine can be used to model a real machine \cite{Turing}. This is not a tautology. The laws governing the real machine are encoded in the \textit{transition function} of the Turing machine. We applied this notion to transfer entropy (TE), a measures for ``information transfer'' between source data and destination data \cite{Lizier2010}. It is also capable of capturing ``true'' causal relations (``true'' in an interventional sense \cite{Pearl}). If we interpret the source data as the input for a Turing machine and the destination data as the output, the transition function should (also) encode causality as captured by TE.
 
 We will start with a short recap of the relevant concepts of Information Theory and transfer entropy. It is then shown that the probability mass function (PMF) of the source data is transformed into the PMF of the destination data occurs via a set of linear transformations. The resulting tensors are used to proof that the Data Processing Inequality or DPI \cite{ThomasCover} is valid for transfer entropy. It is also shown that in well defined cases a bivariate approach suffices to infer the causal structure of a complex system. We end this letter with an experiment to illustrate that our approach is indeed capable of capturing nonlinear relations.

Information theory was introduced in 1948 by C. Shannon \cite{Shannon}. It relates two data sets $x$ and $y$. The data are indexed realizations of quantized random variables representing discrete-time stationary ergodic Markov processes $X$ and $Y$ respectively. If there is dependency between the two messages, information is shared between them. The data are ordered sets of symbols from finite alphabets. In this letter we will use three alphabets: $\mathcal{X}\! =\! \{ \chi_1, \chi_2,\cdots, \chi_{\vert \mathcal{X} \vert} \}$, $\mathcal{Y}\! =\! \{ \psi_1, \psi_2,\cdots, \psi_{\vert \mathcal{Y} \vert} \}$, and $\mathcal{Z}\! =\! \{ \zeta_1, \zeta_2,\cdots, \zeta_{\vert \mathcal{Z} \vert} \}$. The random variable $X$ is associated with the alphabet $\mathcal{X}$, $Y$ with $\mathcal{Y}$, and $Z$ with $\mathcal{Z}$ respectively.

Mutual information (MI) is a measure of the information shared between two time series
\begin{equation} \label{eq:MI}
	I(X;Y) = \sum_{x,y} p(x,y) \log_2 \left[ \frac{p(y|x)}{p(y)} \right].
\end{equation}
It is nonnegative and symmetric in $X$ and $Y$. The information sharing results from data transmission over a \textit{communication channel} (or channel in short). Source data is transmitted, destination data is received. In a channel every input alphabet symbol has it's own input ``socket''. Likewise, every output alphabet symbol has it's own output socket. Data is transmitted one symbol at a time. The input symbol is fed to the related input socket. The channel transforms the input symbol into an output symbol in a probabilistic fashion and makes it available on the associated output socket. The simplest type of channel is the noisy discrete memoryless communication channel (DMC). In a memoryless channel the output ($y_t$) only depends on the input ($x_t$) and not on the past inputs or outputs: $p(y_t \vert x_t, x_{t-1}, y_{t-1})\! =\! p(y_t \vert x_t)$. A memoryless channel embodies the Markov property. The maximum rate with which information can be transmitted over a channel is called the channel capacity $C_{XY}\! =\! \max_{p(x)} \left[ I(X;Y) \right]$. This is achieved for a so called channel achieving input distribution.

In a noisy channel the output depends on the input and another random variable representing perturbations, i.e., noise. Transmission of data over a discrete memoryless communication channel transforms the probability mass function of the input into the PMF of the output via a linear transformation represented by a probability transition matrix \cite{ThomasCover}. This probability transition matrix fully characterizes the DMC. Instead of matrix and vector notation we use index notation. Index $i$ is associated with $x$, index $j$ with $y$, and index $k$ with $z$ respectively. The $j^{th}$ element of the PMF $\textbf{p}(y)$ equals $p(y\! =\! \psi_{j})$. Because every random variable has it's own alphabet letter associated with it this can be written as $p(\psi_{j})$, or even as $p^j$. Using the Einstein summation convention where we sum over double indices, the transmission of $x$ over a noisy channel resulting in $y$ equals
\begin{equation} \label{eq:TransitionTensor}
 	p^j=p^iA^{j}_i.
\end{equation}
 The row stochastic probability transition matrix elements $A^j_i\! =\! p(\psi_j\vert \chi_i)$ represent the elements of the probability transition tensor $\mathsf{A}$ \cite{Prob-Tensor}. In this letter the placement of the indices is used as a mnemonic device. The subscript or covariant index indicates over which alphabet element we have to condition. The superscript or contravariant index indicates which alphabet element is conditioned. It follows directly from Eq.(\ref{eq:TransitionTensor}) that the input distribution can be reconstructed from the output distribution: $p^{j} A^{\ddagger i}_{j}\! =\! p^{i}$, with $A^{\ddagger i}_{j}\! =\! p(\chi_i \vert \psi_j)$. We call this reversal in analysis direction the ${\ddagger}$ operation. If the directed graph \small $X\! \rightarrow\! Y$ \normalsize represents the transmission of data from $X$ to $Y$ with the associated tensor $A^j_i$, the ${\ddagger}$-operation associates $A^{\ddagger i}_{j}$ with \small $X\! \leftarrow^{\ddagger}\! Y$\normalsize.
 
Because mutual information is a function of $\mathsf{A}$ and the input PMF, we write it as \small$I(X,Y)\! :=\! f(\mathsf{A},\star)$\normalsize. The $\star$ indicates that apart from $\mathsf{A}$ there is another input. As such MI might not be the best measure to indicate the underlying structure for systems of which the structure is independent from the input. In contrast, the earlier mentioned channel capacity only depends on the elements of the probability transition tensor \cite{Muroga}. We indicate the channel capacity with the equivalent lower case Greek letter: \small $C_{XY}(\mathsf{A})\! :=\! \bm{\alpha}$, $C_{XY}(\mathsf{B})\! :=\! \bm{\beta}$, and $C_{XY}(\mathsf{C})\! :=\! \bm{\gamma}$\normalsize.

 To understand the usefulness of the tensor formalism we will perform a thought experiment using a simple system consisting of the three random variables $X$, $Y$, and $Z$. We assume that the bivariate relations have the following associated tensors \small $\mathsf{A}\!:\! X\!\rightarrow \! Y$, $\mathsf{B}\!:\! Y\!\rightarrow \! Z$, and $\mathsf{C}\!:\! X\!\rightarrow \! Z$\normalsize. The aim is to determine the true structure: (1) The chain \small $X\! \rightarrow\! Y\! \rightarrow\!Z$\normalsize. (2) The fork \small $X\!\rightarrow \! Y$, $X\!\rightarrow \! Z$\normalsize. (3) The triangle itself. To be able to analyze this graph we need to introduce two concepts \cite{Spirtes2000}: (1) The causal Markov condition. (2) The faithfulness assumption. The Causal Markov Condition states that a process is independent of its non-effects, given its direct causes, i.e., parents. A directed graph is said to be faithful to the underlying probability distributions if the independence relations that follow from the graph are the exact same independence relations that follow from the underlying probability distributions. 
 
Assuming faithfulness and applicability of the causal Markov condition, let's consider the chain. Because it is a straightforward exercise we leave it to the reader to confirm that for the chain we have $C^{k}_i\!=\! A^{j}_i  B^{k}_j$. If we assume that the actual structure is the fork, which can be interpreted as a chain thanks to the $\ddagger$ operation, we get $B^{k}_j\! = \! A^{\ddagger i}_j  C^{k}_i$. From these expressions it follows that we can not distinguish a chain from a fork when $A^{\ddagger i}_{j'}  A^{j}_i \! = \! \delta_{j'j}$ and $A^{j}_{i'} A^{\ddagger i}_{j}  \! = \! \delta_{i'i}$. The Kronecker delta $\delta_{i'i}$ is defined as: $\delta_{i'i} \! =\! 0$ if $i' \! \neq \! i$ and $\delta_{i'i} \!= \! 1$ if $i' \! = \! i$. In this case $\mathsf{A}$ represents a noiseless DMC, i.e., the probability mass function of $y$ is a permutation of the PMF of $x$.

Instead of checking both assumptions we only need to perform one check if we use the DPI. This inequality states that processing of data can never increase the amount of information. For the chain this means that \small$I(X;Z)\! \leq\! \min [I(X;Y),I(Y;Z)]$\normalsize. Only in the absence of noise there is equality. Because the channel capacity is the maximal achievable mutual information for a specific channel, the DPI implies that $\bm{\gamma} \! \leq \! \min [\bm{\alpha} , \bm{\beta} ]$. If $\bm{\gamma} \! < \! \bm{\beta} $, the real structure \textbf{could} be a chain and we have to verify this by using the ``tensor check''. In the case $\bm{\beta} \! < \! \bm{\gamma} $ the real structure \textbf{could} be a fork and we have to check for that. Please note that the tensor expressions are necessary but not sufficient conditions to decide if a relation is false or not. We will discuss the second condition later in this letter. We can not decide between a chain or a fork when $\bm{\gamma} \! =\!\bm{\beta} $.
 
 All this is of course also applicable to time delayed mutual information. Schreiber however showed that time delayed MI is not always capable of determining the correct relation \cite{Schreiber}. Transfer entropy 
\begin{equation} \label{eq:TE}
\small{
TE_{X\rightarrow Y} = \sum_{\textbf{x}^-,y,\textbf{y}^-} p(\textbf{x}^-,y,\textbf{y}^-) \log_2 \left[\frac{p(y|\textbf{x}^-,\textbf{y}^-)}{p(y|\textbf{y}^-)}\right]
}
\end{equation} \normalsize
outperforms time delayed mutual. It is assumed that $Y$ is a Markov process of order $\ell\! \geq\! 1$. With output $y\! =\! y_t$, the relevant past vector of $y$, $\textbf{y}^-\!=\! (y_{t-1}, s, y_{t-\ell})$ and the input vector $\textbf{x}^-\!=\! (x_{t-\tau}, s, x_{t-\tau-m})$ with $m\! \geq\! 0$ and $\tau\! \geq\! 0$. Assuming that there is a finite interaction delay (delay from now on) $\tau$, it is proved that this modified TE is maximal for the real delay \cite{Wibral}. The alphabet for the input vector is $\mathcal{X}^m$, the $m$-ary Cartesian power of the input alphabet $\mathcal{X}$. Likewise, the alphabet for the relevant past vector is $\mathcal{Y}^\ell$, the $\ell$-ary Cartesian power of the output alphabet $\mathcal{Y}$.

From now on we will use the convention that the index $g$ is associated with the relevant past vector of $y$ and $h$ is associated with the relevant past vector of $z$. Transfer entropy can be associated with communication channels. We start with conditioning the MI from Eq.(\ref{eq:MI}) on the event $\textbf{y}^-\!=\! \psi^-_g$
\begin{equation} \label{eq:MI_S}
\small{
  I(X;Y|\psi^-_g) = \sum_{\textbf{x}^-,y} p(\textbf{x}^-,y|\psi^-_g) \log_2 \left[ \frac{p(y|\textbf{x}^-,\psi^-_g)}{p(y|\psi^-_g)} \right].
 }
\end{equation} \normalsize
Because $\textbf{x}^-$ and $\textbf{y}^-$ are the only parents of the output $y$, it follows from the causal Markov condition that the associated channel is memoryless. The conditioned MI quantifies the amount of information that is transmitted over the $g^{th}$ subchannel. Transfer entropy of Eq.(\ref{eq:TE}) can now be expressed as 
\begin{equation} \label{eq:TE_Multi}
	TE_{X\rightarrow Y} = \sum_g p(\psi^-_g) I(X;Y|\psi^-_g).
\end{equation}
Transfer entropy is the result of transmission of data over an \textit{inverse multiplexer}. Let's envision the two time series as data on two parallel vertical tapes. Our inverse multiplexer aligns the tapes by shifting the source data according to the interaction time delay $\tau$. The cell of the input tape containing $x(t-\tau)$ is positioned next to the empty cell of the output tape that will contain $y(t)$. Next it chooses a transmission channel based on the value of the relevant past vector $\textbf{y}^-$. The input vector $\textbf{x}^-$ is fed to an input socket based on the value of the input vector. The channel transforms the input in a probabilistic fashion. The output is written in the appropriate cell of the output tape. To be able to distinguish the input vector index from the output vector index, we indicate an element of the input vector based on $x$ with the index $\hat{i}$. The index $\hat{j}$ is associated with the input vector based on $y$. With \small $p^{j}_{g}\! =\! p(\psi_{j}\vert \psi^{-}_{g})${\normalsize,} $p^{\hat{i}}_{g}\! =\! p(\chi^{-}_{\hat{i}}\vert \psi^{-}_{g})$\normalsize, and \small $A^{j}_{g'\hat{i}}\! =\! p(\psi_j|\chi^{-}_{\hat{i}},\psi^{-}_{g'})$ \normalsize the linear transformation associated with the $g^{th}$ channel is 
\begin{equation} \label{eq:XY}
	p^{j}_{g} = \delta_{g'g} \, p^{\hat{i}}_{g} A^{j}_{g'\hat{i}}.
\end{equation}
The delay for all the subchannels is $\tau_{xy}$. Transfer Entropy is a function of the input PMF and the tensor $\mathsf{A}$. Using the same shorthand notation as for mutual information we define: \small $TE_{X \rightarrow Y} := f (\mathsf{A},\star)$, $TE_{Y \rightarrow Z} := f (\mathsf{B},\star)$, and $TE_{X \rightarrow Z} := f (\mathsf{C},\star)$\normalsize. 
 
We now perform the same thought experiment as previously. First assume that the structure is a chain. Additional to Eq.(\ref{eq:XY}) we have two additional linear transformations
\begin{subequations} 
\begin{align}
	p^{k}_{h} &= \delta_{h'h} \, p^{\hat{j}}_{h} B^{k}_{h'\hat{j}}, \label{eq:YZ}\\
	p^{k}_{h} &= \delta_{h'h} \, p^{\hat{i'}}_{h} C^{k}_{h'\hat{i'}}. \label{eq:XZ}
\end{align}
\end{subequations}
Because \small $\psi^-_{\hat{j}}\! \subset \! \{ \psi_{j},\psi^-_{g} \}$ {\normalsize or} $\{ \psi_{j},\psi^-_{g} \}\! \subset \! \psi^-_{\hat{j}}$ \normalsize we can enlarge either \small $\psi^-_{g}$ {\normalsize or} $\psi^-_{\hat{j}}$ \normalsize so that \small $\{ \psi_{j},\psi^-_{g} \}\! = \! \psi^-_{\hat{j}}$\normalsize. Due to the causal Markov condition this does not impact the end result, so we can replace $j$ by $\hat{j}$. The next step is to condition all sides of Eq.(\ref{eq:XY}) on $\zeta^-_h$ and all sides of Eq.(\ref{eq:YZ}) and Eq.(\ref{eq:XZ}) on $\psi^-_g$. Again thanks to the causal Markov condition we can assume that the cardinality of the input vector for the transformation for \small$X\! \rightarrow\! Z$ \normalsize equals the cardinality of the input vector for the transformation \small$X\! \rightarrow\! Y $\normalsize, i.e., $\hat{i'}\! =\! \hat{i}$. Because we set \small $\{ \psi_{j},\psi^-_{g} \}\! = \! \psi^-_{\hat{j}}$ \normalsize we have \small $B^{k}_{gh\hat{j}}=B^{k}_{h\hat{j}}$\normalsize. The reader can confirm that the causal Markov condition implies that \small $A^{\hat{j}	}_{gh\hat{i}}=A^{\hat{j}}_{g\hat{i}}$\normalsize. Combining the three conditioned equations finally gives us
\begin{equation} \label{eq:XY_conditioned}
	C^{k}_{\hat{i}gh} = A^{\hat{j}}_{g\hat{i}} B^{k}_{h\hat{j}}.
\end{equation}
When we ``sum out'' index $g$ by multiplying both sides with \small $\delta_{i'i}  \delta_{h'h}  p^g_{h'\hat{i'}}$ \normalsize we get Eq.(\ref{eq:GT1}). We repeat these steps assuming that the fork is the true structure. This gives us two expressions for the tensors of the false relations in terms of the tensors of the true relations for both a chain and a fork:
\begin{subequations} \label{eq:GroundTruth}
\begin{align}
C_{h'\hat{i}}^{k}  &= \delta_{h'h} \, \bar{A}^{\hat{j}}_{h'\hat{i}} B_{h\hat{j}}^{k} \text{, with } \bar{A}^{\hat{j}}_{h'\hat{i}} := \delta_{\hat{i}'\hat{i}} \, p^g_{h'\hat{i'}}A_{g\hat{i}}^{\hat{j}}, \label{eq:GT1}\\
B_{h\hat{j}}^{k}  &= \delta_{h'h} \, \bar{A}^{\ddagger \hat{i}}_{h\hat{j}} C_{h'\hat{i}}^{k}\text{, with } \bar{A}^{\ddagger \hat{i}}_{h\hat{j}} := \delta_{\hat{j}'\hat{j}} \, p^g_{h\hat{j'}}A_{g\hat{j}}^{\ddagger \hat{i}}. \label{eq:GT2}
\end{align}
\end{subequations}
Only when \small $\delta_{h'h}  \bar{A}^{\ddagger \hat{i}}_{h\hat{j}'} \bar{A}^{\hat{j}}_{h'\hat{i}} \! = \! \delta_{h'h} \delta_{\hat{j}'\hat{j}}$ {\normalsize and} $\delta_{h'h} \bar{A}^{\hat{j}}_{h'\hat{i}'} \bar{A}^{\ddagger \hat{i}}_{h\hat{j}} \! = \! \delta_{h'h} \delta_{\hat{i}'\hat{i}} $ \normalsize can we not distinguish a chain from a fork. The reader can confirm that $\delta_{h'h} \delta_{\hat{i}'\hat{i}}$ behaves like the identity matrix for every $h$, i.e., it represents a noiseless transmission. When we use the DPI for transfer entropy it follows that this is the case when noise is absent in the relation \small $X \! \rightarrow \! Y$ \normalsize or in the relation \small $Y \! \rightarrow \! Z$ \normalsize or noise is absent in both relations. 

The DPI is a consequence of Eq.(\ref{eq:GT1}). The $h^{th}$ subchannel of the inverse multiplexer of the chain consists itself of a chain of two channels represented by the tensors $\bar{\mathsf{A}}$ and $\mathsf{B}$ with fixed $h$. For this subchannel the DPI is valid: \small $f(\mathsf{C}_h,\star) \leq \min [f(\bar{\mathsf{A}}_h,\star),f(\mathsf{B}_h,\star)]$\normalsize. Transfer entropy is the weighted sum of the TE per subchannel (Eq.(\ref{eq:TE_Multi})). From this it follows \small $f(\mathsf{C},\star) \! \leq \! \min [f(\bar{\mathsf{A}},\star), f(\mathsf{B},\star)]$\normalsize, i.e., the Data Processing Inequality for transfer entropy. The tensor $\bar{\mathsf{A}}_h$ is the result of two cascaded channels represented by $\mathsf{A}_h$ and a tensor with elements \small $p^g_{\hat{i}h}$\normalsize. In this case the DPI leads to \small $f(\bar{\mathsf{A}},\star) \! \leq f(\mathsf{A},\star)$\normalsize. Combining these inequalities we find that for the chain \small$X\! \rightarrow\! Y\! \rightarrow\!Z$\normalsize
\begin{equation} \label{eq:DPI_TE}
	TE_{X\rightarrow Z} \! \leq \min \left[TE_{X\rightarrow Y}, \! TE_{Y\rightarrow Z} \right]. 
\end{equation}
In conjunction with the tensor equations of Eq.(\ref{eq:GroundTruth}) we need to take the delays into account to determine whether a relation is true or false. We posit that interaction delays in a chain are additive. This also applies to a fork because the $\ddagger$-operation is a \textit{time reversal operation}: $\tau^{\ddagger}\! =\! -\tau$. The fork \small$X\! \rightarrow\! Y$, $X\! \rightarrow\! Z$ \normalsize is equivalent to the chain \small$Y\! \rightarrow^{\ddagger}\! X\! \rightarrow\! Z$\normalsize. The total delay for this equivalent chain is $\tau_{yz}=-\tau_{yx}+\tau_{xz}$. The fork is also equivalent to the chain \small$Y\! \leftarrow\! X\! \leftarrow^{\ddagger}\! Z$\normalsize, so \small$\tau_{zy}=-\tau_{zx}+\tau_{xy}$\normalsize. Of these only the relations with a nonnegative total delay could represent physical processes. The proof of additivity is not in scope of this letter, the DPI however makes it plausible. If the total optimal delay in a chain differs from the sum of the individual optimal delays, the TE of at least one individual relation is not maximized. This lowers the upper boundary as given by Eq.(\ref{eq:DPI_TE}). 

To determine when the bivariate approach can not be used we investigated the \textbf{v-structure} \small$X\! \rightarrow Z\! \leftarrow\! Y$\normalsize. Due to the multivariate relation \small$\mathsf{D}\! :\! \{ X,Z\}\! \rightarrow \!Y$ \normalsize there is the additional linear transformation 
\begin{equation} \label{eq:Vstructure}
p^k_{h} = \delta_{h'h} \, p^{\hat{i''} \hat{j'}}_h  D^{k}_{h'\hat{i''}\hat{j'}}. 
\end{equation}
Under the assumption that $\hat{i''}\! = \! \hat{i'}$ and $\hat{j'}\! = \! \hat{j}$ and using the fact that \small $p^{\hat{i}\hat{j}}_{h}\! =\! \delta_{h'h} \, \delta_{\hat{i'}\hat{i}} \, p^{\hat{i}}_{h} p^{\hat{j}}_{h\hat{i'}}$\normalsize we get the following two relations relating $\mathsf{D}$ to both $\mathsf{B}$ and $\mathsf{C}$:
\begin{subequations} \label{eq:vstructure}
\begin{align}
C_{h\hat{i}}^{k} &= \delta_{h'h} \, \delta_{\hat{i'}\hat{i}} \, p^{\hat{j}}_{h\hat{i'}}  D^{k}_{h'\hat{i}\hat{j}},\\
B_{h\hat{j}}^{k} &= \delta_{h'h} \, \delta_{\hat{j'}\hat{j}} \, p^{\hat{i}}_{h\hat{j'}}  D^{k}_{h'\hat{i}\hat{j}}. 
\end{align}
\end{subequations} 
Let's assume that $\hat{i}\! \leq\! N$ and $\hat{j}\! \leq\! M$. In the bivariate approach we want to determine the tensor $\mathsf{D}$ using the bivariate measurements. The reader can confirm that this is only possible in the case $N,M \in \{1,2\}$. If there are 2 or more indirect paths between two nodes the bivariate analysis can, in theory, not be used.

\begin{figure} [t]
\centering
\includegraphics[scale=0.55]{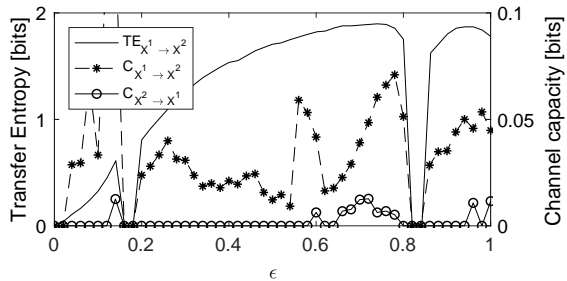}
 \caption{\label{Ulam_Map} The channel capacity for the relations $X^{1} \rightarrow X^{2}$ and $X^{2} \rightarrow X^{1}$ as function of the coupling strength $\epsilon$. Dots: channel capacity for quantized data. Line: transfer entropy as determined by Schreiber.}
\end{figure}

We finalize this letter with an experiment to illustrate that nonlinear behavior is indeed captured by measuring the probability transition tensors and calculating the channel capacities. We use the one-dimensional lattice of unidirectional coupled maps $ x^{m}_{n+1}\! =\! f \left( \epsilon x^{m-1}_{n}\! +\! (1-\epsilon) x^{m}_{n}\right)$. Information can only be transferred from $X^{m-1}$ to $X^{m}$. The Ulam map with $f(x) = 2−x^2$ is interesting because there are two regions ($\epsilon\! \approx\! 0.18$, $\epsilon\! \approx\! 0.82$) where no information is shared between maps \cite{Schreiber}. We used the following quantization scheme: if $x_{n-1} \geq x_{n} < x_{n+1}$ or $x_{n-1} < x_{n} \geq x_{n+1}$ then $x'_n:=1$ otherwise $x'_n:=0$. Furthermore we chose $\ell=m=1$ (see Eq.(\ref{eq:TE})). Instead of maximizing TE we maximized the channel capacity to determine the optimal delay. In the case of none or weak autoregressive data we use the upper boundary
\small
\begin{equation} \label{eq:Causal_Cp}
  \max_{p(x)} \left[ TE_{X\rightarrow Y} \right] \leq \sum_{g} p(\psi^-_g) C_{XY|\psi^-_g}.
\end{equation}\normalsize
The relations for the set $\{X^1,X^2\}$ were measured with significance level 0.01. The delays were varied between 1 and 20. The Channel capacity is maximal for an delay of 1 sample. As can be seen in figure \ref{Ulam_Map}, the \textit{structure} is identical to the one as determined by Schreiber
\\

To conclude, we have shown that we are capable of determining the causal structure as far as encoded in the probability mass functions of quantized time series. Instead of computing transfer entropy, we determine probability transition tensors that transform source data into destination data. These were used to show that in specific cases bivariate analysis suffices to distinguish false relations from true relations. We also used it to derive the Data Processing Inequality for transfer entropy. Our approach is only applicable to noisy data. No assumptions were made about the cardinality of the alphabets. This implies that there must be an equivalent approach for non-quantized data.

I would like to thank Errol Zalmijn for introducing me to the wonderful topic of transfer entropy and Marcel Brunt for helping me to implement these principles in Matlab. Also thanks to Hans Onvlee, S. Kolumban, Rui M. Castro and T. Heskes for their comments on earlier versions of the manuscript. This work performed under the auspices of ASML PI System Diagnostics.


\begin{thebibliography}{}

\bibitem{Schreiber}
Thomas Schreiber.
\newblock Measuring information transfer.
\newblock {\em Phys. Rev. Lett.}, 85:461--464, Jul 2000.

\bibitem{Runge}
Jakob Runge, Jobst Heitzig, Vladimir Petoukhov, and J\"urgen Kurths.
\newblock Escaping the curse of dimensionality in estimating multivariate
  transfer entropy.
\newblock {\em Phys. Rev. Lett.}, 108:258701, Jun 2012.

\bibitem{Wollstadt}
Michael Wibral, Patricia Wollstadt, Ulrich Meyer, Nicolae Pampu, Viola
  Priesemann, and Raul Vicente.
\newblock Revisiting wiener's principle of causality - interaction-delay
  reconstruction using transfer entropy and multivariate analysis on
 delay-weighted graphs.
\newblock In {\em Annual International Conference of the {IEEE} Engineering in
 Medicine and Biology Society, {EMBC} 2012, San Diego, CA, USA, August 28 -
 September 1, 2012}, pages 3676--3679, 2012.

\bibitem{Turing}
A.~M. Turing.
\newblock {On Computable Numbers, with an Application to the
 Entscheidungsproblem}.
\newblock {\em Proceedings of the London Mathematical Society},
 s2-42(1):230--265, 01 1937.

\bibitem{Lizier2010}
J.~T. Lizier and M.~Prokopenko.
\newblock Differentiating information transfer and causal effect.
\newblock {\em The European Physical Journal B}, 73(4):605--615, Feb 2010.

\bibitem{Pearl}
Judea Pearl.
\newblock {\em Causality: Models, Reasoning and Inference}.
\newblock Cambridge University Press, New York, NY, USA, 2nd edition, 2009.

\bibitem{Shannon}
C.~E. Shannon.
\newblock A mathematical theory of communication.
\newblock {\em Bell System Technical Journal}, 27(3):379--423.

\bibitem{ThomasCover}
Thomas~M. Cover and Joy~A. Thomas.
\newblock {\em {Elements of Information Theory}}.
\newblock Wiley-Interscience, New York, NY, USA, 1991.

\bibitem{Prob-Tensor}
Wen Li and Michael K.~Ng.
\newblock On the limiting probability distribution of a transition probability
 tensor.
\newblock {\em Linear and Multilinear Algebra}, 62, 03 2014.

\bibitem{Muroga}
S.~{Muroga}.
\newblock {On the Capacity of a Discrete Channel.}
\newblock {\em Journal of the Physical Society of Japan}, 8:484--494, July
 1953.

\bibitem{Spirtes2000}
Peter Spirtes, Clark Glymour, Scheines N., and Richard.
\newblock {\em Causation, Prediction, and Search}.
\newblock Mit Press: Cambridge, 2000.

\bibitem{Wibral}
Michael Wibral, Nicolae Pampu, Viola Priesemann, Felix Siebenh\"uhner, Hannes
 Seiwert, Michael Lindner, Joseph~T. Lizier, and Raul Vicente.
\newblock Measuring information-transfer delays.
\newblock {\em PloS one}, 2013.

\end{thebibliography}
\end{document}